\documentclass[10pt,letterpaper,twocolumn,prl,showpacs]{revtex4-1}

\usepackage[english]{babel}

\usepackage{amsmath,amssymb}
\usepackage[dvips]{graphicx,color}
\usepackage{hyperref}

\begin{document}

\title{Experimental observation of coherent cavity soliton frequency combs in silica microspheres}

\author{Karen E. Webb$^*$, Miro Erkintalo, St\'ephane Coen, and Stuart G. Murdoch}

\affiliation{The Dodd-Walls Centre for Photonic and Quantum Technologies, Department of Physics, The University of Auckland, Private Bag 92019, Auckland 1142, New Zealand \\
$^*$Corresponding author: kweb034@aucklanduni.ac.nz }

\begin{abstract}
We report on the experimental observation of coherent cavity soliton frequency combs in silica microspheres. Specifically, we demonstrate that careful alignment of the microsphere relative to the coupling fiber taper allows for the suppression of higher-order spatial modes, reducing mode interactions and enabling soliton formation. Our measurements show that the temporal cavity solitons have sub-100-fs durations, exhibit considerable Raman self-frequency shift, and generally come in groups of three or four, occasionally with equidistant spacing in the time domain. RF amplitude noise measurements and spectral interferometry confirm the high coherence of the observed soliton frequency combs, and numerical simulations show good agreement with experiments.
\end{abstract}

\maketitle

\noindent The generation of optical frequency combs in microresonators was first demonstrated in 2007 \cite{delhaye07}, and has since attracted considerable interest \cite{kippenberg11,delhaye11,okawachi11}. Such microresonator-based combs are an attractive alternative to commercially available mode-locked-laser based systems \cite{udem02}, as in addition to their small footprint and high power efficiency, they can possess a very large mode spacing and power per mode, which may be useful for applications such as telecommunications \cite{pfeifle14}. However, most of the initially observed microresonator-based combs were incoherent and noisy, which made them unsuitable for many applications. A solution to this problem has only recently emerged, with the demonstration that microresonators can support temporal cavity solitons (CSs): pulses of light that can persistently propagate around the cavity without changes in their shape or energy. First observed and studied in passive fiber ring resonators \cite{leo10}, such CSs correspond to steady-state solutions of the Lugiato-Lefever equation that describes the universal dynamics of Kerr nonlinear cavities \cite{coen13a}. Significantly, they manifest themselves as highly coherent optical frequency combs in the spectral domain \cite{herr14}. 

The first observations of temporal CSs in microresonators were made in a magnesium fluoride device \cite{herr14}, and similar observations have subsequently been reported in several other platforms, including silica wedge resonators \cite{yi15,yi16} and silicon nitride resonators \cite{brasch16, joshi16}. However, so far CS frequency combs have only been realized in microresonators that are fabricated using comparatively complex or time-consuming techniques. This may be because resonators made using simpler methods often suffer from a large number of spatial modes that hinder soliton formation. Specifically, the interaction between different spatial modes can lead to avoided mode crossings which strongly alter the local dispersion of the pumped mode family, thus preventing soliton formation \cite{herr14b}. Although several techniques that mitigate the effect of mode interactions have been demonstrated \cite{savchenkov05,murugan11,ding12,kordts16,huang16}, they have not, to the best of our knowledge, been applied to achieve soliton formation in resonators that are simple, cheap, and fast to fabricate.

\looseness=-1 In this Letter, we demonstrate the formation of CS frequency combs in one of the simplest microresonator geometries: silica microspheres \cite{braginsky89,agha09}. In particular, we show that, by changing the angle between a silica microsphere and the fiber taper used to couple light into it, we can suppress higher-order spatial modes to the extent that coherent CS frequency combs can be excited via a ``power-kick'' method \cite{kordts16,brasch16,yi16,brasch16arxiv}. In addition to RF amplitude noise measurements, we verify that these structures are highly coherent by using spectral interferometry \cite{webb16}. We also use numerical simulations to support our experimental findings and to explain why only certain soliton configurations can be experimentally excited. We believe our work will improve the accessibility of microresonator-based soliton frequency combs, thus facilitating broader experimental studies.

\begin{figure}
\centering
\includegraphics[width=\linewidth]{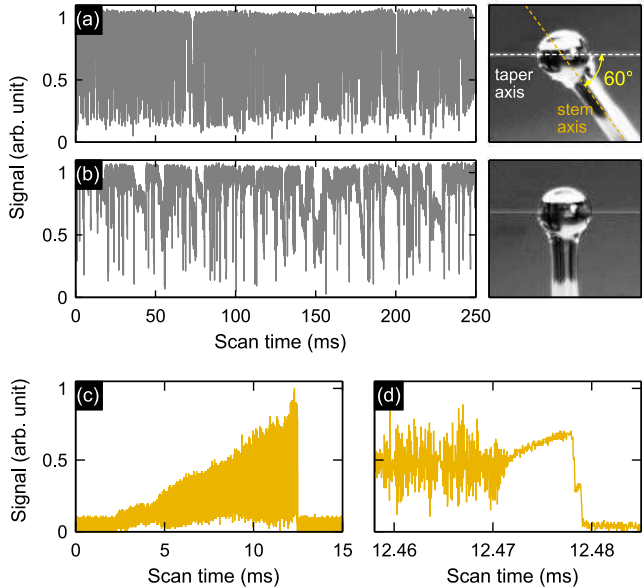}
\caption{Scan of microsphere mode spectrum across one FSR when the sphere stem is at an angle of (a) 60~degrees, and (b) 90~degrees relative to the fiber taper. Corresponding images of the sphere coupled to the taper are shown to the right. (c) Scan of typical cavity resonance when sphere angle and coupling has been optimized. (d) Close-up of the end of the same resonance as in (c), showing low-noise steps which indicate the presence of CSs. Scans in (c) and (d) are measured at the taper output through an offset filter.}
\label{fig:modes}
\end{figure}
The silica microspheres are fabricated by melting the end of a standard $125~\mu$m diameter single-mode optical fiber in a fiber fusion splicer: the melted tip of the fiber pulls itself into a near-perfect sphere due to surface tension \cite{spillane02}. The resulting microspheres typically have diameters of about $250~\mu$m, free-spectral range $\textrm{FSR}\approx260$~GHz, and finesse $\mathcal{F}\approx5\cdot10^4$ ($Q\approx3.7\cdot10^7$). To excite frequency combs, we use a setup similar to that presented in \cite{webb16}. The pump is derived from an external cavity diode laser at 1550~nm, which is amplified with an erbium-doped fiber amplifier and then filtered with a 3~nm bandpass filter to produce an 80~mW continuous wave field. This field is then launched into a fiber taper that has a waist diameter of~$\approx1~\mu$m, which is coupled to the silica microsphere. The taper output is directly monitored with an optical spectrum analyzer and a photodetector. We also monitor the output through an offset filter on a second photodetector. This offset filter passes all wavelengths above 1555~nm, and the resulting signal is, to a good approximation, proportional to the average power contained in the intracavity frequency comb.

We first discuss the microsphere mode spectrum. Specifically, due to the size and geometry of the spheres, they support a large number of spatial modes \cite{schiller93}. This leads to the presence of numerous avoided mode crossings, which are known to be detrimental to soliton formation \cite{herr14b}. However, the number of higher-order modes (and hence mode crossings) can be reduced by adjusting the angle of the fiber stem relative to the taper axis \cite{murugan11,kalkman06}. This is highlighted in Figs.~\ref{fig:modes}(a) and (b). Here we show the mode spectrum obtained by scanning the laser frequency across one FSR of the microsphere for two different angles between the sphere and the taper. For consistency, we allow the sphere to touch the taper for each scan. At an angle of 60~degrees, shown in Fig.~\ref{fig:modes}(a), the mode spectrum appears very dense. As the angle is increased, however, we find that the mode spectrum becomes considerably less dense, thus reducing the chance of encountering avoided mode crossings. The cleanest mode spectrum is observed when the taper is perfectly perpendicular to the fiber stem. We find that we can achieve additional suppression of higher-order modes by touching the taper to the sphere a small vertical distance away from the equator. Figure~\ref{fig:modes}(b) shows the mode spectrum measured for optimal alignment, and we can indeed see how the number of modes is substantially reduced. Significantly, under these coupling conditions, we can routinely see signatures of CS formation. Indeed, Fig.~\ref{fig:modes}(c) shows a scan across a cavity resonance  measured through the offset filter, while Fig.~\ref{fig:modes}(d) shows a section of the same scan at the end of the resonance. We see clean steps evidencing the formation of CSs \cite{herr14}. We emphasize that, for non-optimal coupling, we have not observed any such steps.

\begin{figure}
\centering
\includegraphics[width=\linewidth]{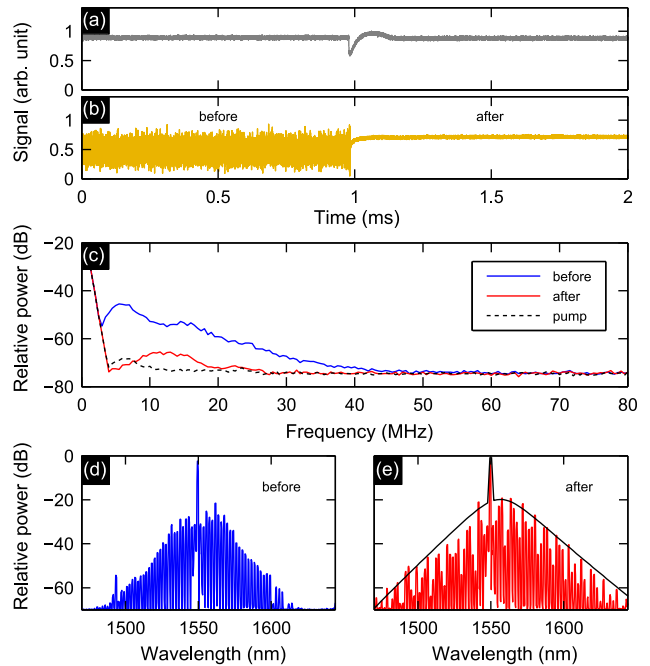}
\caption{(a,b)~Power kick used to excite CSs; (a) shows the pump power, and (b)~shows the taper output through an offset filter. Kick pulse is applied at $\approx1$~ms. (c)~RF amplitude noise measurements of the combs before (blue solid line) and after (red solid line) the power kick, respectively. Dashed line indicates amplitude noise of pump. (d),~(e)~Experimentally measured spectra of frequency combs before and after the power kick, respectively. Black curve in (e) represents simulated single soliton envelope with $\Delta=13.5$. }
\label{fig:kick}
\end{figure}
To study in detail the characteristics of the CS frequency combs, we must first lock the detuning of the resonator inside the narrow range that supports them. Because the microspheres have a small volume, thermal effects are comparatively strong \cite{ilchenko92,carmon04,karpovarxiv}, and so we are unable to scan the laser frequency at sufficient speed to reach the soliton states. Instead we employ a power kicking technique \cite{kordts16,brasch16,yi16,brasch16arxiv}. Specifically, we first manually increase the laser wavelength until it reaches a detuning that gives rise to an incoherent ``modulation instability'' comb, to which we thermally lock \cite{carmon04}. We then use an amplitude modulator located immediately after the pump laser to quickly drop the pump power for a short time, which causes the resonator to rapidly cool and the resonances to blue shift, effectively scanning the pump towards the CS region \cite{yi16}. Figures~\ref{fig:kick}(a) and (b) show, respectively, the pump power and the average comb power (measured through the offset filter) during the process. At $t\approx1$~ms, a pulse from an arbitrary waveform generator is applied to the modulator to produce an exponential power kick pulse, with $\textrm{FWHM}=20~\mu$s and a depth of 30\%. After the kick, the comb transitions to a state that corresponds to a resonance step, where it is again thermally locked. The transition is clearly visible in the reduction of noise in the photodetector signal measured after the offset filter [Fig.~\ref{fig:kick}(b)]. This noise reduction is highlighted in more quantitative terms in Fig.~\ref{fig:kick}(c), where we show the measured RF amplitude noise for the comb states before and after the power kick. The transition is also clearly observed in the measured spectra. Figure~\ref{fig:kick}(d) shows the spectrum before the power kick, and it can be seen to have the smoothness characteristic of an unstable modulation instability comb \cite{erkintalo14}. In contrast, the spectrum of the comb after the kick [Fig.~\ref{fig:kick}(e)] shows strong spectral interference fringes, clearly revealing a change in the comb state.

\begin{figure}[t]
\centering
\includegraphics[width=\linewidth]{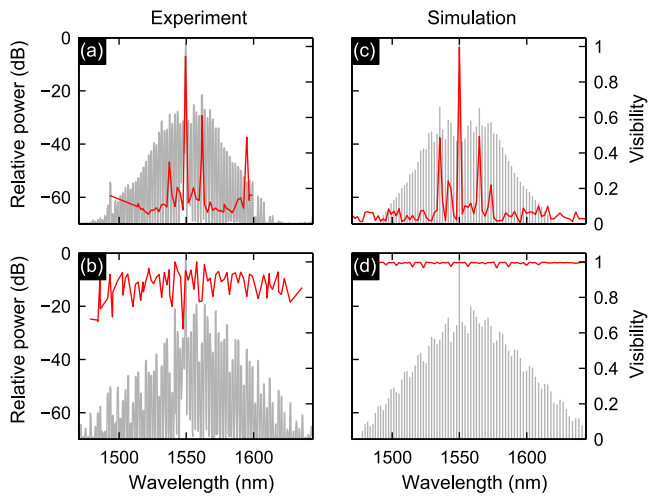}
\caption{Red solid lines show (a)--(b) experimentally measured and (c)--(d) numerically simulated visibility of (a), (c) an unstable modulation instability comb and (b), (d) a coherent state corresponding to three cavity solitons (right axis). Interferometer delay was set to 100~m in all cases. Gray lines show measured and simulated comb spectra for reference (left axis). Cavity soliton state shown in (b) was reached by power kicking from the unstable comb state in (a). Numerical simulations use detuning values (c) $\Delta=0.5$ and (d) $\Delta=13.5$.}
\label{fig:coh1}
\end{figure}
The observed reduction in RF noise, and the spectral characteristics after the power kick, provide strong evidence of the formation of coherent CSs. In particular, the spectral fringes seen in Fig.~\ref{fig:kick}(e) can be explained as a result of the interference between multiple CSs that are temporally distributed around the resonator \cite{herr14,brasch16}. By taking the Fourier transform of these fringes, we can find the corresponding angular positions of the CSs \cite{brasch16}. We find that the spectrum in Fig.~\ref{fig:kick}(e) is indicative of three CSs, with angles of $0^{\circ}$, $95.4^{\circ}$, and $194.4^{\circ}$. In Fig.~\ref{fig:kick}(e), we also plot the numerically simulated spectrum corresponding to a single CS in order to highlight the sech$^2$ envelope (see below for simulation details), and find that it agrees well with the envelope of the experimental spectrum. We note that the comb is red-detuned relative to the pump by approximately 10~nm. This is likely due to the Raman response of silica causing a strong self-frequency shift, which has been reported to affect frequency combs corresponding to ultrashort solitons \cite{milian15,karpov16}.

We have further confirmed the coherence of the CS comb state by measuring the line-by-line coherence using the spectral interferometry technique presented in \cite{webb16}. Specifically, by interfering resonator outputs separated by a fixed number of roundtrips, spectral fringes can be observed; the visibility of these fringes is equal to the magnitude of the complex degree of first-order coherence \cite{erkintalo14}. To experimentally measure the visibility, we use a fiber-based delayed Michelson interferometer as described in \cite{webb16}. For all measurements, the interferometer delay is set to 100~m, corresponding to approximately 15~photon lifetimes, which is more than sufficient to draw conclusions concerning the coherence of the comb state \cite{erkintalo14}. Red lines in Figs.~\ref{fig:coh1}(a)~and~(b) show the experimentally measured spectral visibilities of the combs previously shown in Figs.~\ref{fig:kick}(d) and (e), respectively. (Experimentally measured spectra are also shown in gray for clarity.) Before the power kick [Fig.~\ref{fig:coh1}(a)], only the pump and a pair of symmetrically detuned sidebands show high or moderate visibility. These features are characteristic of the unstable modulation instability regime identified in \cite{erkintalo14}. In contrast, the comb after the kick shows high visibility across the full spectrum [Fig.~\ref{fig:coh1}(b)], as expected for a soliton comb.

To verify our experimental results, we have also carried out numerical simulations using an extended Lugiato-Lefever equation that incorporates the Raman response of silica \cite{coen13a,karpov16}:
\begin{equation}
\begin{split}
t_\textrm{R}\frac{\partial E(t,\tau)}{\partial t}=&\left[-\alpha-i\delta_0-\frac{iL\beta_2}{2}\frac{\partial^2}{\partial\tau^2}\right]E+\sqrt{\theta}E_\textrm{in}\\
&+i\gamma L\left[R(\tau)\ast|E(\tau)|^2\right]E.
\end{split}
\label{lle}
\end{equation}
Here, $E(t,\tau)$ is the intracavity field, $t$ and $\tau$ are the slow and fast times, respectively,  $t_\textrm{R}=1/\textrm{FSR}$ is the cavity roundtrip time, $\alpha=\pi/\mathcal{F}$ describes the total cavity losses, $\delta_0$ is the cavity detuning, $L$ is the cavity length, $\beta_2$ is the group-velocity dispersion, $E_\textrm{in}$ is the pump field, and $\theta$ is the coupling power transmission coefficient. The convolution term describes the nonlinear contributions, where $\gamma$ is the nonlinear interaction coefficient, and $R(\tau)=(1-f_\textrm{R})\delta(\tau)+f_\textrm{R}h_\textrm{R}(\tau)$ is the nonlinear response function including the Kerr nonlinearity and Raman scattering, where $f_\textrm{R}=0.18$ is the Raman fraction of silica, and $h_\textrm{R}(\tau)$ is the Raman response function obtained from the multiple vibrational mode model \cite{hollenbeck02}. The group-velocity dispersion and nonlinear coefficients are estimated as $\beta_2=-7$~ps$^2$/km and $\gamma=2$~W$^{-1}$km$^{-1}$, respectively, using analytic expressions for a microsphere \cite{schiller93}. The normalized detuning $\Delta=\delta_0/\alpha$ is chosen for each comb to best match the experimental results (see captions of Figs.~\ref{fig:kick} and \ref{fig:coh1}). We integrate \eqref{lle} using the split-step Fourier method, starting with an initial condition consisting of the upper continuous-wave state solution with quantum noise \cite{erkintalo14}. The numerically simulated visibility and spectra of the unstable modulation instability and CS combs are shown in Figs.~\ref{fig:coh1}(c) and (d), respectively. They are in good qualitative agreement with the experimental results in Figs.~\ref{fig:coh1}(a) and (b).

\begin{figure}
\centering
\includegraphics[width=\linewidth]{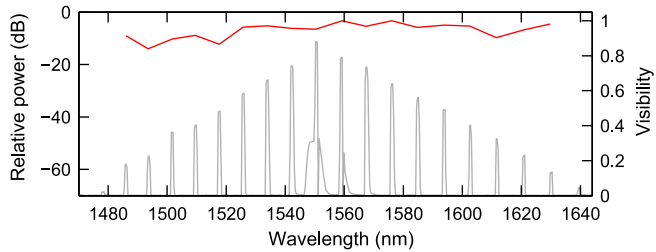}
\caption{Red solid line shows experimentally measured visibility of a coherent state corresponding to four CSs equally spaced around the cavity (right axis). Interferometer delay was set to 100~m. Gray line shows measured comb spectrum for reference (left axis). Cavity soliton state was reached by power kicking from an unstable comb.}
\label{fig:coh2}
\end{figure}

Our observations of coherent CS combs are not unique to the particular microsphere studied. Indeed, we have performed additional experiments for several microspheres (which differ slightly due to variation in the fabrication process), and systematically found soliton combs when higher-order modes are appropriately suppressed. Experimentally, we find that the CS combs always correspond to states with either three or four solitons in the resonator. Since the number of solitons generated by each power kick is expected to vary stochastically, we believe that only the states with three or four solitons are able to maintain thermal locking in our system (after the power kick). We suspect that this also explains why only approximately one out of ten power kick attempts are successful; the remaining attempts likely result in a thermally unstable configuration of solitons. Numerical simulations (not shown here) corroborate these interpretations, by showing that approximately 10\% of trials result in the generation of an appropriate number of solitons. Although the arrangement of the solitons can vary from realization to realization, we occasionally find that the solitons are exactly equidistant. The measured spectrum and coherence of such a state is shown in Fig.~\ref{fig:coh2}. This state was reached in a different microsphere for a reduced pump power of 40~mW. (Power kicking from an unstable modulation instability comb was again used to reach the coherent state.) As can be seen, the comb exhibits a high degree of coherence across the spectrum, and has a mode spacing of four FSR, thus strongly alluding to the presence of four CSs exactly equally spaced around the resonator. We note that similar configurations of equally spaced solitons have very recently been observed in other resonator platforms \cite{brasch16,joshi16,lamb16cleo}. We speculate that this behaviour is due to the the presence of weak mode crossing perturbations, which lead to oscillatory features in the time domain that can trap the CSs to precise positions \cite{herr14b,jang15,wang16cleo,lamb16cleo}. Finally, we remark in closing that attempts to excite single solitons through the backward scanning technique \cite{karpovarxiv} have been unsuccessful due to the small thermal locking range of this system.

In conclusion, we have observed coherent frequency combs corresponding to temporal CSs in optical microspheres. By suppressing higher-order modes that would otherwise prevent soliton formation, we have been able to excite CSs in this system. Measurements of the spectral coherence and amplitude noise of the combs confirm their high degree of coherence. We expect that the demonstration of CS formation in such a simple platform will facilitate the experimental study of coherent microresonator frequency combs.

\textbf{Funding.} Marsden Fund and Rutherford Discovery Fellowships of the Royal Society of New Zealand.


\newcommand{\enquote}[1]{``#1''}

\end{document}